\documentclass[11pt]{article}

\usepackage[margin=1in]{geometry}
\usepackage[T1]{fontenc}
\usepackage[utf8]{inputenc}
\usepackage{lmodern}
\usepackage{microtype}
\usepackage{booktabs}
\usepackage{array}
\usepackage{tabularx}
\usepackage{listings}
\usepackage{xcolor}
\usepackage{amsmath,amssymb}
\usepackage{url}
\usepackage[hidelinks]{hyperref}

\title{Solving VeriContest with a Lean-Backed Rust Verifier}

\author{Traian Serbanuta \and Jun Xu \and Andrei Stefanescu \and Cosmin Radoi}

\date{Asymptotic \\[0.5em] October 2026}

\begin{document}
\maketitle

\begin{abstract}
VeriContest is a benchmark of 1007 competitive-programming problems in Rust, each with a Verus specification, a judge-accepted solution, and a Verus proof. Its authors report that proof generation is the bottleneck for frontier models: given the specification and the code, the best model produces an accepted Verus proof for 13.95\,\% of the problems on the first attempt. We report on solving the same proof-generation task with Rust-Prover, a verifier for Rust backed by Lean~4. The Verus specification and the Rust code are restated and translated into Lean, each specification becomes a theorem, and agents prove the theorems with Lean's kernel as the final check. All 1325 theorems of all 1007 problems were proved by Claude Opus 5.5, in about 35 hours of wall clock, at a median of 3.4 minutes and \$1.20 per proof, with 71\,\% of them accepted at the first official check. The restated specifications were checked against the benchmark's test suites, and reviewed where no suite applies. None was wrong or weakened. The translated Lean programs were run on 19,555 of the benchmark's test cases and produced the same output as the Rust programs on every one. Across five Claude and four GPT models at five reasoning-effort settings, every current frontier model proves nearly all of a ten-theorem sample at every setting, and more effort mostly raises the cost rather than the number of proofs. Sonnet 5.5 at low effort proves all of the 50 hardest theorems, a GPT-6 Luna-first run with Opus 5.5 rescues proves the whole benchmark again at a third of the recorded cost, and Claude Haiku 5.5 proves that run's ten most expensive theorems at its top effort settings. We also rerun the benchmark's own Verus protocol with Claude Opus 5.5 on the 50 problems with the longest reference proofs. Opus 5.5 alone fails to prove one of them.
\end{abstract}

\section{Introduction}

Much of the infrastructure that runs blockchains and the services around them is written in Rust. We have built Rust-Prover, a verifier for Rust backed by Lean~4~\cite{lean4}, to verify such code as it ships. Rust-Prover works in three steps. Agents write a specification from a description of the intended behaviour. The code and the specification are translated into Lean, where each specification becomes a theorem. Automated tactics and agents prove the theorems, and Lean's kernel checks the proofs. We have used it on Solana crates since June 2026.

This paper reports on running Rust-Prover against VeriContest~\cite{vericontest2026}, a public benchmark of about a thousand competitive-programming problems in Rust with expert-reviewed Verus~\cite{verus2023} specifications and proofs. We take the benchmark's proof-generation task, in which a model receives the specification and the code and must produce a proof, and solve it with Lean proofs instead of Verus proofs. Our contributions are the following.

\begin{itemize}
  \item All 1325 theorems of all 1007 problems were proved and re-checked in a clean clone, with no axioms beyond Lean's three standard ones (Section~\ref{sec:proofs}).
  \item The restated specifications were validated against the benchmark's test suites, and by review where no suite applies (Section~\ref{sec:spec}). The translated programs were validated on 19,555 executed cases from the benchmark's test suites (Section~\ref{sec:code}).
  \item A study of five Claude and four GPT models at five reasoning-effort settings, and of cheap models on the hardest theorems and on the whole benchmark (Section~\ref{sec:models}).
\end{itemize}

\section{VeriContest}

VeriContest is a benchmark for verifiable code generation in Rust, published in May 2026. Its main set has 946 problems, 690 from LeetCode and 256 from Codeforces. Each problem comes with the original natural-language description, a Verus specification reviewed by experts, a Rust solution accepted by the online judge, a Verus proof that the solution meets the specification, and two test suites, one of correct cases and one of mutated cases that a correct specification must reject. The repository also ships 61 extended problems, verified the same way but left out of the paper's evaluation because their outputs need a problem-specific judge. The extended problems have no test suites. In total there are 1007 problems. The counts in this paper refer to the September 2026 state of the benchmark, pinned at commit b9657ed. The repository has since been restructured and now counts 1006 problems.

The benchmark defines four tasks for a model.

\begin{itemize}
  \item \textbf{SpecGen.} Read the description and write the specification. Graded mainly by running the specification against both test suites.
  \item \textbf{CodeGen.} Write the program from the description, the specification or both. Graded on the tests.
  \item \textbf{ProofGen.} Given the specification and the code, make Verus accept them. That means adding proof annotations to the code, such as loop invariants, assertions, termination measures and calls to helper lemmas, until the solver can check the specification. The specification and the code may not be changed.
  \item \textbf{End2End.} Do all three from the description alone.
\end{itemize}

The authors evaluated ten frontier models. The highest scores were GPT-5.5's, shown in Table~\ref{tab:paper}. Code generation is almost solved. Proof generation is the bottleneck.

\begin{table}[h]
\centering
\begin{tabularx}{\textwidth}{Xr}
\toprule
Task & pass@1 \\
\midrule
CodeGen (description to code) & 92.18\,\% \\
SpecGen (description to spec) & 48.31\,\% \\
ProofGen (spec + code to Verus proof) & 13.95\,\% \\
End2End & 5.29\,\% \\
\bottomrule
\end{tabularx}
\caption{GPT-5.5 on VeriContest, as reported by the benchmark's authors. pass@1 is the share of problems solved on the first attempt. The model is asked once and that one output is checked.}
\label{tab:paper}
\end{table}

\section{Rust-Prover on VeriContest}

We challenge Rust-Prover with ProofGen. For each problem it gets the Verus specification, the Rust code, and the problem's English description. It has to produce a Lean proof that the code meets the specification. The benchmark's own proof annotations, such as loop invariants, are not translated, and the agents never see them.

\subsection{Step 1. Restate the specification}

The Verus specification is first restated as a Rust-Prover spec. A transpiler does this where it can, and an agent writes the spec for the rest, following a porting guide.

Take problem 7, Reverse Integer, as an example. The benchmark's code is a simple accumulator loop (abridged):

\begin{lstlisting}
pub fn reverse(x: i32) -> i32 {
    if x == i32::MIN { return 0; }
    let neg = x < 0;
    let mut cur: i32 = if neg { -x } else { x };
    let mut res: i32 = 0;
    while cur != 0 {
        // omitted: res = res * 10 + cur % 10 with checked arithmetic, return 0 on overflow
        cur = cur / 10;
    }
    if neg { -res } else { res }
}
\end{lstlisting}

The benchmark's specification requires that the function return the digits of a signed 32-bit integer in reverse, and 0 exactly when the reversed number does not fit. Restated for Rust-Prover, it reads as follows (abridged).

\begin{lstlisting}
#[pure]
pub fn reverse_spec_helper(x: Nat, acc: Nat) -> Nat {
    decreases(x);
    if x == 0.to_nat() { acc } else {
        let new_acc = acc * 10.to_nat() + x % 10.to_nat();
        if new_acc > i32::MAX.to_int().to_nat() { 0.to_nat() }
        else { reverse_spec_helper(x / 10.to_nat(), new_acc) }
    }
}

#[pure]
pub fn reverse_spec(x: Int) -> Int {
    if x >= 0.to_int() {
        reverse_spec_helper(x.to_nat(), 0.to_nat()).to_int()
    } else {
        -reverse_spec_helper((-x).to_nat(), 0.to_nat()).to_int()
    }
}

#[spec]
pub fn reverse_main_spec(x: i32) -> i32 {
    requires(i32::MIN <= x);
    requires(x <= i32::MAX);
    let out = reverse(x);
    ensures(out.to_int() == reverse_spec(x.to_int()));
    out
}
\end{lstlisting}

The restated specification is plain Rust, in the style of our Sui Prover. In the example above, the spec is \texttt{reverse\_main\_spec}, marked with \texttt{\#[spec]}. It calls \texttt{reverse} and states in its \texttt{ensures} that the output must equal \texttt{reverse\_spec(x)}. The expected value is defined by \texttt{reverse\_spec} and its helper, marked with \texttt{\#[pure]}. They reverse the digits over unbounded naturals, give 0 when the result does not fit an \texttt{i32}, and handle the sign. \texttt{decreases} says the recursion terminates.

\subsection{Step 2. Translate the crate into Lean}

The crate, spec included, is translated into Lean by our forks of Charon~\cite{charon2025} and Aeneas~\cite{aeneas2022}. Aeneas turns safe Rust into pure Lean functions. A \texttt{\&mut} argument becomes a value that goes in and a value that comes out, which is sound because Rust's ownership rules guarantee nobody else can touch that value while the reference is alive. A loop becomes a recursive body function. Every arithmetic operation returns a \texttt{Result}, so an overflow that would panic in Rust becomes a \texttt{fail} in Lean. The output is a Lean program with the same meaning as the crate.

The spec's helper functions are handled differently. Instead of going through Aeneas, the \texttt{\#[pure]} macro rewrites their Rust bodies directly into Lean definitions, and after the translation our driver inserts those definitions into the Lean program. So \texttt{reverse\_spec\_helper} arrives in Lean as an ordinary recursive definition, one a proof can unfold and do induction on.

\subsection{Step 3. Turn the spec into a theorem}

At this point the Lean program contains the implementation, the helper functions, and the spec. The spec is then turned into a theorem statement by \texttt{\#derive\_spec}, a Lean command we wrote. For \texttt{reverse\_main\_spec} the statement fits on one line:

\begin{lstlisting}
theorem reverse_main_spec_correct (x : I32)
    (hreq0 : decide (core.num.I32.MIN ≤ x) = true)
    (hreq1 : decide (x ≤ core.num.I32.MAX) = true) :
    ∃ r, reverse_main_spec x = ok r
\end{lstlisting}

The theorem says that for any \texttt{x} in range, \texttt{reverse\_main\_spec x} returns a result. A failed \texttt{ensures} would make it return an error instead, so returning a result means the postcondition held. Inputs whose reversal does not fit need no special treatment. Both the code and \texttt{reverse\_spec} give 0 for them, so the postcondition also holds.

Each \texttt{\#[spec]} function becomes one theorem. Many problems have more than one, since a postcondition with a quantifier is stated as its own \texttt{\#[spec]} function. The 1007 problems give 1325 theorems in total.

\subsection{Step 4. Prove the theorems}

Proving the theorems happens in three stages, each picking up what the previous one left.

\paragraph{Automation proves what it can.} A fixed sequence of Lean tactics runs on every theorem first. When it fails it leaves behind the exact goal it could not close.

\paragraph{Agents write the rest.} The agents work in a repository that holds, for each problem, the Rust code, the problem's description, the restated specification, and the Lean translation ending in the theorem with its proof left as \texttt{sorry}. An agent takes one theorem at a time and works out everything the proof needs. It has three hours and 40 iterations. An iteration is an official check: it submits the current proof for acceptance and returns the errors to revise against. Between checks the agent can run Lean as often as it likes on scratch copies, and most of the revision happens there. It may not touch the code or the statement. If it believes the theorem is false, it releases it with its reason.

\paragraph{Every proof passes the same check.} The agent may only fill in the proof. The theorem statement and the machine-generated Lean around it must be byte-identical to the pipeline's output, and Lean must report no axioms beyond its three standard ones (\texttt{propext}, \texttt{Classical.choice}, \texttt{Quot.sound}). The axiom check is also what forbids \texttt{sorry}, \texttt{admit}, \texttt{native\_decide} and \texttt{bv\_decide}. Nothing the theorem depends on may use them. Every accepted proof is re-checked from scratch in a clean clone, and a proof that fails the re-check counts as unproved.

When nothing closes a theorem, it stays open, with the agent's notes recorded for the next attempt.

For \texttt{reverse}, most of the work is a lemma about the loop, proved by induction with this invariant, quoted from the accepted proof:

\begin{lstlisting}
inv := fun s => 0 ≤ s.1.val ∧ 0 ≤ s.2.val ∧
  p212.reverse_spec_helper_pure s.1.val.toNat s.2.val.toNat = R
\end{lstlisting}

The state \texttt{s} is the pair of \texttt{cur} and \texttt{res} from the code. The invariant says both stay non-negative and that continuing from this state ends at the expected answer. The rest of the proof handles the code outside the loop and applies the lemma. Opus 5.5 wrote the proof in three official checks and about two minutes, adding about 150 lines of Lean. The benchmark's own Verus proof states essentially the same invariant, with one extra bound, \texttt{res <= i32::MAX}, that the Lean proof did not need. The agent had never seen that proof. It arrived at the invariant on its own.

\section{Results}

\subsection{Specification translation}
\label{sec:spec}

All 1007 specifications were translated, 768 by the transpiler and 239 by agents running on Claude Opus 5.5. To validate them, we ran each translated spec, which is executable Rust, against the benchmark's test suites. It must accept every correct case and reject every mutated case. This validated 658 of the 1007 specs. The other 349 could not be checked by the test suites, and an agent reviewed them against the Verus original under a written protocol.

\begin{itemize}
  \item 134 have a quantifier the check cannot enumerate and 44 are extended problems with no suite. We sampled 20 specs from each group and reviewed them.
  \item 169 had a suite that could not finish and were all reviewed one by one. 23 of those were sampled for a second review that is not given the first verdicts.
  \item For the remaining 2, special-judge problems whose suites accept several answers, we checked every answer our spec accepted against the benchmark's own specification, and all were valid.
\end{itemize}

The reviews found no wrong or weakened spec.

\subsection{Code translation}
\label{sec:code}

All 1007 programs translate to Lean, and every theorem statement is covered.

To validate the translation, we executed the generated Lean definition of the program on the benchmark's own test cases and compared the results with the Rust program's. Of the 114 problems with test suites the harness could try, 105 ran to completion: 19,555 executed cases, sampled from the suites, every one agreeing with the Rust program and none disagreeing. The other 9 problems could not be run, because of timeouts or signatures the execution harness does not support.

For 27 problems, the specification or the code could not be translated as it stood, and we made a specialization to get the translation through. The benchmark's code or a spec helper was re-encoded in a form the toolchain or the spec crate could handle, with evidence that the meaning is unchanged. Appendix~\ref{app:specializations} lists them.

\subsection{Proof generation}
\label{sec:proofs}

All 1325 theorems were proved and passed the re-check. That is every theorem of every one of the 1007 problems. The proofs were written by Claude Opus 5.5 over about 35 hours of wall clock, with three agents working in parallel.

Table~\ref{tab:perproof} gives the figures per proof.

\begin{table}[h]
\centering
\begin{tabularx}{\textwidth}{Xrrrr}
\toprule
Per proof & Median & Mean & 90th percentile & Max \\
\midrule
Time, claim to submit & 3.4 min & 4.3 min & 8.4 min & 69.7 min \\
Iterations (budget 40) & 1 & 1.46 & 3 & 7 \\
Cost & \$1.20 & \$1.35 & \$2.40 & \$5.67 \\
\bottomrule
\end{tabularx}
\caption{All 1325 proofs.}
\label{tab:perproof}
\end{table}

A typical proof took an agent about three minutes from claiming the theorem to submitting the proof, and nine in ten took under nine minutes. The slowest took just over an hour.

Most proofs were accepted at the first official check. 944 of the 1325 were accepted on the first check, 238 on the second, 89 on the third, and the remaining 54 within seven. No proof came near the budget of 40. A proof cost about a dollar at Claude Code's list prices, and the most expensive about six. The whole run cost \$1,792 and delivered 38 proofs per hour.

We gave each problem a rough difficulty estimate, a unitless score where higher is harder. Table~\ref{tab:difficulty} covers 1266 of the 1325 proofs: 30 problems have no estimate and 29 have an estimate of zero, and both groups are left out. Harder problems took longer and cost more, but the gap is modest. The hardest class took about three and a half times the time and three times the money of the easiest, and no more iterations. Difficulty changed the price of a proof, not whether we got one.

\begin{table}[h]
\centering
\begin{tabularx}{\textwidth}{Xrrrr}
\toprule
Difficulty & Proofs & Median time & Mean iterations & Median cost \\
\midrule
0 to 50 & 405 & 2.4 min & 1.64 & \$0.87 \\
50 to 100 & 253 & 3.2 min & 1.36 & \$1.20 \\
100 to 200 & 315 & 4.0 min & 1.42 & \$1.39 \\
200 to 400 & 220 & 5.0 min & 1.36 & \$1.54 \\
400 and up & 73 & 8.6 min & 1.19 & \$2.63 \\
\bottomrule
\end{tabularx}
\caption{The main run's proofs by difficulty estimate.}
\label{tab:difficulty}
\end{table}

\subsection{Models and reasoning effort}
\label{sec:models}

The run used Claude Opus 5.5 at one reasoning-effort setting. We wanted to know how much that choice mattered, so we repeated a small part of the run as a sweep of five Claude models and four GPT models at five effort settings each.

We took 10 theorems from the run, spread across its range of difficulty. Opus had spent between 1.1 and 9.7 minutes on them. Each model at each setting attempted each theorem once, alone in a fresh environment with only the theorem and the instructions, and with a budget of 10 iterations and three hours. The GPT models ran through Codex. We stopped running Haiku 4.5 on the hardest theorems after it failed the easier ones, so it made 43 attempts in place of 50. Table~\ref{tab:proved} shows how many theorems each model proved and Table~\ref{tab:cost} the cost per proved theorem.

\begin{table}[h]
\centering
\begin{tabularx}{\textwidth}{Xrrrrr}
\toprule
Theorems proved & low & medium & high & xhigh & max \\
\midrule
\multicolumn{6}{l}{\textit{Claude}} \\
Claude Fable 5.1 & 10 of 10 & 10 of 10 & 10 of 10 & 10 of 10 & 10 of 10 \\
Claude Opus 5.5 & 10 of 10 & 10 of 10 & 10 of 10 & 10 of 10 & 10 of 10 \\
Claude Sonnet 5.5 & 9 of 10 & 10 of 10 & 10 of 10 & 10 of 10 & 10 of 10 \\
Claude Haiku 5.5 & 8 of 10 & 10 of 10 & 10 of 10 & 10 of 10 & 10 of 10 \\
Claude Haiku 4.5 & 0 of 9 & 0 of 8 & 0 of 9 & 0 of 8 & 0 of 9 \\
\multicolumn{6}{l}{\textit{GPT}} \\
GPT-6 Astra & 10 of 10 & 10 of 10 & 10 of 10 & 10 of 10 & 10 of 10 \\
GPT-6.1 Sol & 10 of 10 & 10 of 10 & 10 of 10 & 10 of 10 & 10 of 10 \\
GPT-6 Sol & 9 of 10 & 10 of 10 & 10 of 10 & 10 of 10 & 10 of 10 \\
GPT-6 Luna & 0 of 10 & 0 of 10 & 3 of 10 & 8 of 10 & 8 of 10 \\
\bottomrule
\end{tabularx}
\caption{Theorems proved by model and reasoning-effort setting.}
\label{tab:proved}
\end{table}

Fable, Opus, GPT-6 Astra and GPT-6.1 Sol proved every theorem at every setting. Sonnet and GPT-6 Sol proved all but one, each missing it at low effort. Haiku 5.5 missed two at low effort and proved everything from medium up. GPT-6 Luna proved nothing below high effort and 8 of 10 at the top two settings. Haiku 4.5 proved none. For the models that succeeded, the median number of iterations per proof was one at every setting, except GPT-6 Astra and GPT-6 Sol at low effort, which needed two and three, and GPT-6 Astra at high, at one and a half. More effort bought more proofs only for Luna and Haiku 5.5, which went from 8 of 10 at low effort to all 10 at medium.

\begin{table}[h]
\centering
\begin{tabularx}{\textwidth}{Xrrrrr}
\toprule
Cost per proved theorem & low & medium & high & xhigh & max \\
\midrule
\multicolumn{6}{l}{\textit{Claude}} \\
Claude Fable 5.1 & \$3.59 & \$4.71 & \$5.41 & \$7.40 & \$7.97 \\
Claude Opus 5.5 & \$1.20 & \$1.43 & \$1.56 & \$2.22 & \$3.57 \\
Claude Sonnet 5.5 & \$0.95 & \$0.94 & \$0.97 & \$1.40 & \$2.56 \\
Claude Haiku 5.5 & \$0.44 & \$0.48 & \$0.62 & \$0.78 & \$1.04 \\
\multicolumn{6}{l}{\textit{GPT}} \\
GPT-6 Astra & \$2.38 & \$2.21 & \$2.28 & \$2.69 & \$2.99 \\
GPT-6.1 Sol & \$0.39 & \$0.40 & \$0.39 & \$0.44 & \$0.49 \\
GPT-6 Sol & \$1.20 & \$1.00 & \$1.02 & \$1.18 & \$1.32 \\
GPT-6 Luna & -- & -- & \$0.19 & \$0.17 & \$0.22 \\
\bottomrule
\end{tabularx}
\caption{Cost per proved theorem. GPT costs are estimates at API list prices from the tokens reported, since those runs went through Codex.}
\label{tab:cost}
\end{table}

More effort did raise the bill. For the Claude models the highest setting cost 2x to 3x the lowest, and took longer. For the GPT models the rise was a quarter or less. GPT-6.1 Sol was the cheapest at about \$0.40 a proof, Haiku 5.5 next at about \$0.50, Sonnet about a dollar, and Fable the most expensive. Haiku 5.5 at medium effort was the cheapest Claude setting with full coverage, at about half Sonnet's cost. Median times were 4 to 8 minutes a proof for Opus and Sonnet, 5 to 8 for GPT-6 Astra, 7 to 12 for Haiku 5.5, and 17 to 21 for GPT-6 Sol.

Ten theorems is a small sample with few hard ones in it, so we consider these figures as indications.

Can cheap models still do the harder work? The next three experiments each pick up the question the previous one left open.

The first question is whether a cheap setting can prove the hardest theorems. Sonnet 5.5 at low effort was the cheapest Claude setting before Haiku 5.5. We took the 50 theorems the run had spent the most time on, where Opus had taken between 10 and 70 minutes each. We gave each theorem to Sonnet at low effort under the same conditions as before, alone with the theorem and the instructions, 10 iterations and three hours. The instructions had been revised in the meantime with lessons from the run. Table~\ref{tab:sonnet} gives the outcome.

\begin{table}[h]
\centering
\begin{tabularx}{\textwidth}{Xr}
\toprule
Sonnet 5.5 at low effort, 50 hardest theorems & \\
\midrule
Proved & 50 of 50 \\
Accepted on the first iteration & 48 of 50 \\
Median time per proof & 16.3 min \\
Median cost per proof & \$2.17 \\
Total cost & \$118 \\
\bottomrule
\end{tabularx}
\caption{Sonnet 5.5 at low effort on the 50 theorems the run spent the most time on.}
\label{tab:sonnet}
\end{table}

Sonnet proved all fifty, 48 of them at the first iteration. On the same fifty theorems in the run, Opus had a median of 13.8 minutes and \$2.73 a proof, with 44 at the first iteration. Sonnet was cheaper on 38 of the 50 and slower on most, taking about 1.3x as long. The two runs differ in more than the model, so this is not a controlled comparison. It does say that the hardest theorems in the benchmark are within reach of the smaller model at its lowest effort setting.

With the hardest theorems within a cheap setting's reach, the next question is the cost of the whole benchmark. GPT-6 Luna was the cheapest model in the study and the weakest of the frontier models. We ran the whole benchmark again with Luna at xhigh effort trying every theorem first and Opus 5.5 taking only the theorems Luna failed. Each model had 10 iterations and three hours per theorem, and each theorem was worked in an isolated repository with no other proofs to draw on, though Opus started its rescues from Luna's failed attempts. All 1325 theorems were proved, under the same acceptance check but without the clean-clone re-check reported for the main run. Luna proved 1146 of them and Opus proved the other 179. The recorded cost of the whole run was \$643, or \$0.48 per proof. 348 of the 2,198 Luna sessions, about 24 hours of session time, left no usage record, so the true cost is somewhat higher.

Opus alone had recorded \$1,792 for the same 1325 theorems, so the Luna-first run cost 64\,\% less, and all of the saving came from the theorems Luna settled on its own. The theorems that fell through to Opus cost about 9\,\% more than Opus alone, since Luna's failed attempts were added to the bill. The saving comes at the cost of more time. The median proof in this new evaluation took 32 minutes, against 3 in the Opus run; the 32 includes abandoned attempts and operational recovery, so it measures the run more than the model.

That run leaves the last question. Its most expensive theorems still needed Opus, a big model, and we wanted to know whether a cheap model could cover even those. We took the ten theorems the run had spent the most on: the five most expensive that Luna proved alone and the five most expensive of Opus's rescues. We gave each to Claude Haiku 5.5 once per effort setting, from scratch, under the same protocol as the sweep. At xhigh and max effort it proved all ten, rescues included. The ten cost \$34.64 at max, about what the recorded Luna-plus-Opus pipeline had spent on the same ten (\$33.09), and on the rescue theorems alone Haiku 5.5 spent 13\,\% less than Opus had. At low effort it gave up on all ten within about two minutes each, spending \$0.15 in total. Every give-up was on a theorem its higher efforts proved. These are single attempts on ten theorems, so again an indication, but the cheapest Claude model covers the most expensive theorems of the benchmark when told to try hard enough.

\subsection{Comparison with the VeriContest paper}
\label{sec:comparison}

Our results and the results the VeriContest paper reports for frontier models are not strictly comparable, for two reasons.

\begin{enumerate}
  \item The VeriContest paper reports pass@1. Each model is asked once for a proof and that single output is checked. Our agents get up to 40 iterations of checking and revising, with Lean's feedback after each.
  \item The models are different. The highest ProofGen score in the VeriContest paper is GPT-5.5's, and our agents run on Claude Opus 5.5.
\end{enumerate}

So we ran the paper's own ProofGen task ourselves with Claude Opus 5.5, following the setting the benchmark suggests. We allowed 10 iterations and three hours per problem, since no proof in our run needed more than 7 iterations. We ran it on the 50 problems whose reference proofs are the longest in the benchmark, from 527 to 1,777 added lines. Table~\ref{tab:verus} gives the outcome.

\begin{table}[h]
\centering
\small
\begin{tabularx}{\textwidth}{Xrrrrr}
\toprule
Opus 5.5, 50 longest reference proofs & Total & Median & Mean & 90th pct. & Max \\
\midrule
Proved & 49 of 50 & & & & \\
Accepted on the first attempt & 18 of 50 & & & & \\
Time per proof & & 6.1 min & 7.4 min & 12.0 min & 26.7 min \\
Iterations per proof (budget 10) & & 2 & 2.31 & 4.2 & 10 \\
Cost per proof & & \$0.92 & \$1.12 & \$2.05 & \$4.14 \\
\bottomrule
\end{tabularx}
\caption{The benchmark's own Verus protocol, run with Claude Opus 5.5. Time is from the first request to acceptance.}
\label{tab:verus}
\end{table}

Our run proved all 1325 theorems, 71\,\% at the first official check, at a median of 3.4 minutes and \$1.20 per proof. Of the 50 problems here, 47 map to our problem set and carry 65 theorems. Our run proved all 65, 50 of them at the first official check, at a median of 8.8 minutes per theorem. Opus 5.5 under the paper's protocol proved 49 of these 50 problems, 36\,\% on the first attempt, at a median of 6.1 minutes and \$0.92 per proof. There is one difference the numbers do not show. Ours are Lean proofs, mathematical objects that anyone can re-check with nothing but the file. A Verus pass is a solver's verdict, and it cannot be re-checked outside Verus.

The two setups also differ in other ways.

Our agents had more to work with. The rerun was a bare API call given the specification and the code. Our agents had tools, the problem's English description, free Lean runs between official checks, and worked examples in their instructions. In the run they could also draw on other accepted proofs and a similarity index. So 71\,\% accepted at the first official check and 36\,\% on a single API call are not comparable rates.

The rerun had its own advantages. It covers 50 problems, against all 1007 problems and all 1325 theorems in ours. And the benchmark's reference proofs in Verus have been public on GitHub since the paper came out in May 2026, while Opus 5.5 is newer than that, so a model asked for a Verus proof may have seen the published answer. No Lean proof of any of these theorems existed anywhere before we wrote them, so there was no answer to recall and every proof had to be found. What remains is indirect. The Rust code and its Verus proof are public, and a model that studied the code may remember an invariant the Verus proof used. That is a hint at most, not a proof, and it still had to be turned into a Lean proof that the kernel accepts.

\section{Conclusion}

Every theorem of every VeriContest problem now has a Lean proof. Agents wrote the proofs at a median of three minutes and about a dollar each, and Lean's kernel checked every one, with no axioms beyond the standard three. The benchmark's authors found proof generation to be the bottleneck for frontier models. With Lean's feedback to iterate against, it is not.

Three questions remain open. First, we do not yet know how much of this carries over to the crates we verify for customers, where the specifications are written from scratch rather than restated from a benchmark. Second, we have not found the limit of the cheap models: Sonnet 5.5 at low effort proved the hardest fifty theorems, a Luna-first run proved the whole benchmark at a third of the recorded cost, and Haiku 5.5 proved its ten most expensive theorems. Third, a score on a public benchmark may owe something to having seen its answers. The benchmark's Verus proofs are public, so for the Verus rerun this cannot be ruled out. No Lean proof of these theorems existed before we wrote them, so for our run there was no answer to recall, as Section~\ref{sec:comparison} discusses.

\bibliographystyle{plain}
\bibliography{refs}

\appendix

\section{The 27 specializations}
\label{app:specializations}

\begin{table}[h]
\centering
\small
\begin{tabular}{>{\raggedright\arraybackslash}p{1.9cm}>{\raggedright\arraybackslash}p{2.2cm}>{\raggedright\arraybackslash}p{3.7cm}>{\raggedright\arraybackslash}p{2.9cm}>{\raggedright\arraybackslash}p{3.7cm}}
\toprule
Specialization & Problems & What was done & Why & Evidence of unchanged meaning \\
\midrule
Implementation rewritten & 2 (lc3046, lc3233) & A nested \texttt{for} with an early return became a \texttt{while}. \texttt{f64::sqrt} became an integer floor square root. Specs unchanged. & The translation could not join that loop's exits. No floating-point model. & Exhaustive or differential testing against the old code. Review. \\
\addlinespace
Spec helper redefined & 1 (lc1700) & A helper that rotates a queue was redefined as a counting model. Code unchanged. & No list construction in the spec crate. & Equivalence proof by induction. 420,000-case differential test against the Verus helper. The test suites. Review. \\
\addlinespace
Strings as bytes & 17 (ASCII string problems) & \texttt{String} and \texttt{Vec<char>} became \texttt{Vec<u8>}, character literals became bytes, in code and spec. Suite strings are fed in as bytes. & No \texttt{char} type in the spec crate. & Each problem's precondition restricts inputs to ASCII, so nothing is lost. Differential test against a \texttt{Vec<char>} transcription. Review. \\
\addlinespace
Pairs as parallel vectors & 5 (cf437C, cf479C, cf1788C, lc163, lc885) & A \texttt{Vec<(A, B)>} became two vectors of equal length, in code and spec. & No tuple element types in the spec crate. & Bijection argument per problem. Differential test against a tuple transcription. Review. \\
\addlinespace
Hash maps as vectors & 2 (lc421, lc523) & A \texttt{HashSet} and a \texttt{HashMap} became vector lookups with the same contents at every step. Specs unchanged. & Standard hash maps bring opaque hasher axioms into the translation, so no proof over them is kernel-clean. & Same result by construction. Review. \\
\bottomrule
\end{tabular}
\caption{The 27 problems whose code or spec helper was re-encoded to get the translation through.}
\end{table}

\end{document}